\documentclass[a4paper]{spie}  
\usepackage{amsmath,amssymb}
\usepackage{graphicx}
\usepackage[colorlinks=true, allcolors=blue]{hyperref}
\usepackage{enumerate} 

\title{Quality analysis of DCGAN-generated mammography lesions}
\author[a]{Basel Alyafi}
\author[b]{Oliver Diaz}
\author[c]{Joan C Vilanova}
\author[d]{Javier del Riego}
\author[a]{Robert Marti}
\affil[a]{Computer Vision and Robotics Institute, University of Girona, Spain}
\affil[b]{Dept. of Mathematics and Computer Science, University of Barcelona, Spain}
\affil[c]{{Faculty of Medicine, University of Girona, Girona, Spain}}
\affil[d]{Dept. of Radiology, UDIAT Centre Diagnostic, Parc Tauli Hospital Universitari, Institut d'Investigacio Parc Tauli I3PT, Spain}

\begin{document} 
\maketitle
\begin{abstract}
Medical image synthesis has gained a great focus recently, especially after the introduction of Generative Adversarial Networks (GANs). GANs have been used widely to provide anatomically-plausible and diverse samples for augmentation and other applications, including segmentation and super resolution. In our previous work, Deep Convolutional GANs were used to generate synthetic mammogram lesions, masses mainly, that could enhance the classification performance in imbalanced datasets. In this new work, a deeper investigation was carried out to explore other aspects of the generated images evaluation, i.e., realism, feature space distribution, and observers studies. t-Stochastic Neighbor Embedding (t-SNE) was used to reduce the dimensionality of real and fake images to enable 2D visualisations. Additionally, two expert radiologists performed a realism-evaluation study. Visualisations showed that the generated images have a similar feature distribution of the real ones, avoiding outliers. Moreover, Receiver Operating Characteristic (ROC) curve showed that the radiologists could not, in many cases, distinguish between synthetic and real lesions, giving $48\%$ and $61\%$ accuracies in a balanced sample set.
\end{abstract}
\section{INTRODUCTION}
\label{sec:intro}
\indent Generative Adversarial Networks (GANs) \cite{GAN} have been widely used in medical image synthesis for more than two years. This technique has shown promising results in different modalities and for different organs and applications: colour retinal images synthesis \cite{end2end_retinal}, CT lung patches for lesion detection augmentation \cite{liver_aug}, MRI brain lesion removal/addition for segmentation and classification (augmentation) tasks \cite{A2NGAN_brainLesion}, to name a few. However, many of those works were focused solely on improving the performance of a deep-learning-based model (application-driven evaluation). On the other hand, few papers shed a light over visualising the feature space of the generated images and/or reviewing specialists opinion \cite{liver_aug}. In a previous work \cite{DCGAN_mammo_ours}, we used Deep Convolutional GANs \cite{radford_DCGAN} and proved that the generated synthetic images were effective in enhancing the classification process even in unbalanced conditions (where the class under study (positive) was minor to the normal (negative) class). In order to conduct an integrated study and to satisfy the requirements of image synthesis definition (realism and anatomic plausibility conditions) in Ref. \citenum{simul_synth_mia}, different aspects of the evaluation process of the synthetic images should be examined. Here, we give more importance to the nature of the synthetic images (previously-generated masses and microcalcifications), i.e. to visualise the 2D feature space using t-Stochastic Neighbor Embedding (t-SNE) \cite{t_SNE}. Additionally, the professional opinion of two radiologists was provided through a computerised study to analyse how realistic the generated images were in the eyes of specialists.
\section{Materials}
The dataset used in this work is OPTIMAM Mammography Image Database (OMI-DB) \cite{Optimam}. This database includes more than 145,000 cases (over 2.4 million images) and comprises unprocessed and processed digital mammograms from the Breast Screening Programme of the United Kingdom. A subset of this database was obtained comprising over 80,000 cases. In this dataset, there are images from four vendors, however, only images belonging to Hologic Selenia Dimensions (Hologic, Inc; Bedford, Massachusetts, USA) were used in this work. This database has expert annotations identifying the image and any clinical observation. A total of 5,351 mass and microcalcification lesions and 22,000 normal tissue patches were extracted with size $128 \times 128$ pixels after applying histogram normalisation. 

\section{Methodology}
\begin{figure}[h]
\centering
\includegraphics[scale=1]{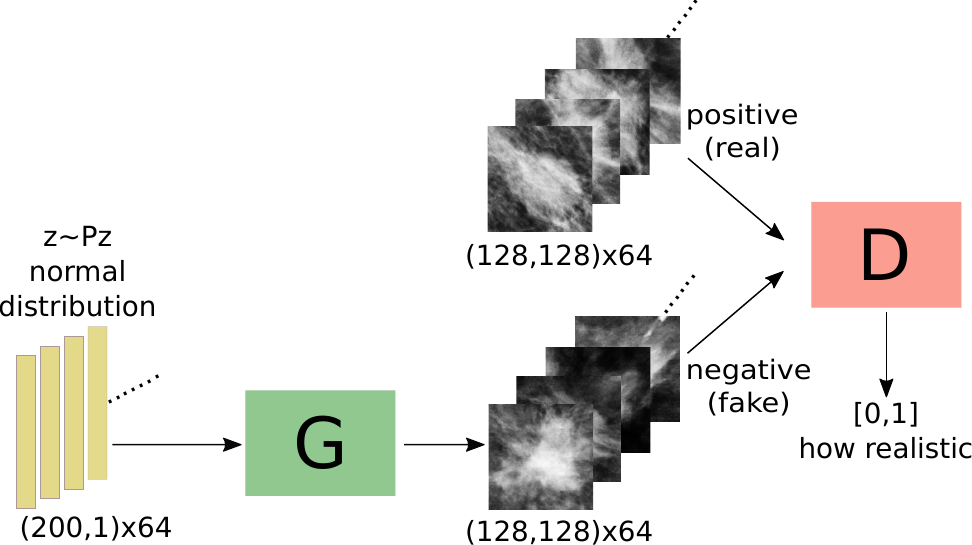}
\caption{Training schematic for DCGAN. The generator (G) takes samples from the normal distribution ($\mu=0,\ \sigma=1$ and maps it to mammographic lesions. The discriminator (D) decides how realistic the input is (the larger the output the higher the realism).}
\label{fig:GAN_training}
\end{figure}

As described earlier in our previous work \cite{DCGAN_mammo_ours}, DCGANs \cite{radford_DCGAN} were used to generate synthetic mammography lesions and microcalcifications. The process of training the DCGAN is explained in the schematic in Fig. \ref{fig:GAN_training}. In this figure, a batch of 64 vectors of length 200 is sampled from the latent space $\mathcal{N}(0,1)$ and input to the generator G which learns how to map them to the distribution of mammographic lesions. Thereafter, the discriminator D learns to distinguish between real and synthetic lesions giving a value between 0 and 1 (zero denotes definitely fake while one denotes definitely real). For more details about DCGAN training techniques used, see Ref. \citenum{DCGAN_mammo_ours}.

\section{Experiments}
To test the realism of the generated images (patches), we projected the patches into a 2D feature space which allows to see the location of each cluster (real mass lesions, fake mass lesions, and real normal tissue). Additionally, a human-observer study was performed with radiologists in order to evaluate the realism of the synthetic images.
\subsection{t-SNE experiment}
\label{sec:tsne_experiment}
t-SNE was used to reduce the dimensionality of the images (patches) from $128 \times 128$ to 2 using 500 patches of each cluster (real mass lesions, synthetic mass lesions, and real normal tissue). The parameters used were 4000 for iterations number and 250 for perplexity.
\subsection{Radiologists study}
\label{sec:radiologist_study}
Two radiologists from two different hospitals in Catalonia (Spain), with 7 and +25 years in breast radiology, participated in a human observer study using a balanced random sample of 150 patches ($128 \times 128$ pixels) containing cancerous mass lesions (75 real and 75 synthetic). For each image, they had 6 options to choose among: extremely, moderately, or slightly confident real; or extremely, moderately, or slightly confident synthetic. These options were then converted to numerical values by assigning the probabilities $\{0.95,\ 0.77,\ 0.59,\ 0.41,\ 0.23,\ 0.05\}$ where the highest value was for extremely confident real. These numbers were then used for calculating the accuracy (using 0.5 as threshold) and for drawing the Receiver Operating Characteristic (ROC) curve.

\section{Results and discussion}
\subsection{t-SNE analysis}
\begin{figure}[h]
\centering
\includegraphics[scale=1]{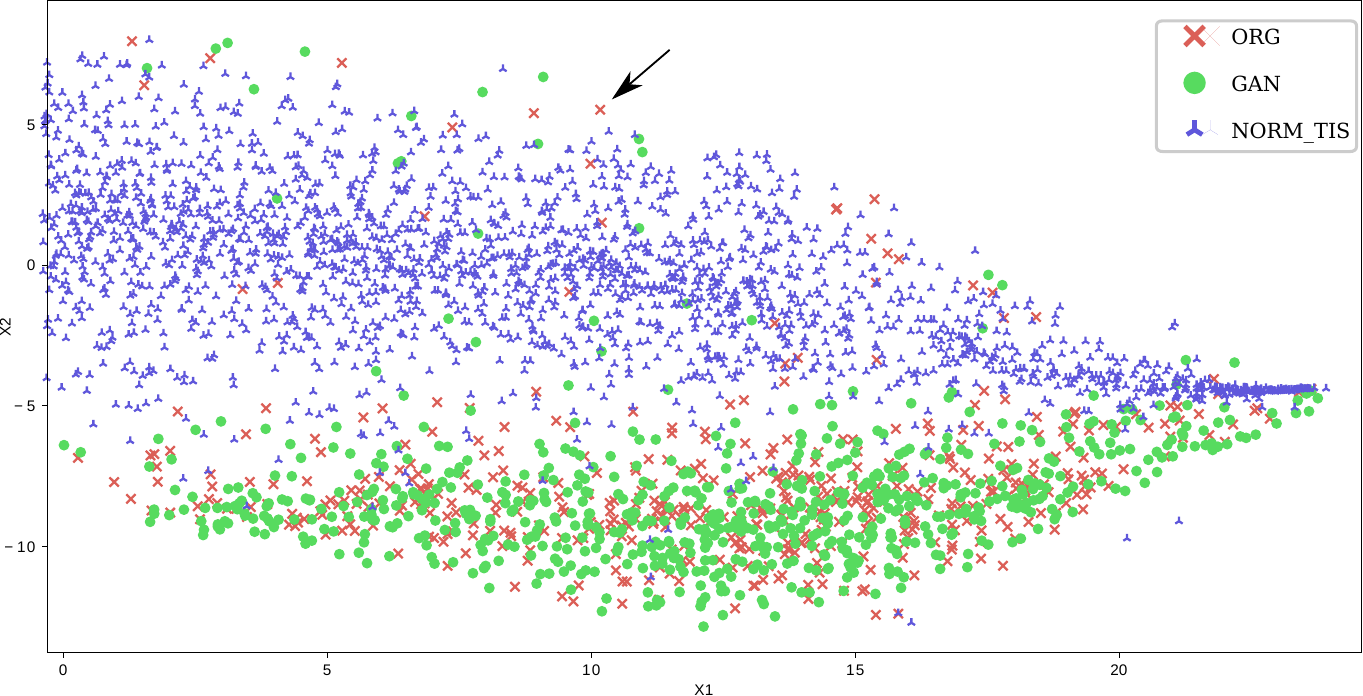}
\caption{2D t-SNE representation of real (red crosses) and fake (green circles) lesions and real normal tissue (purple triangles). The arrow points at an exemplar real lesion outlier.}
\label{fig:tsne_real_fake}
\end{figure}

As a result of the experiment described in Sec. \ref{sec:tsne_experiment}, Fig. \ref{fig:tsne_real_fake} shows the distribution of each of the following: 
i) real lesions (red crosses), ii) synthetic lesions (green circles), and iii) real normal tissue (purple triangles). It is clear from this figure that the distribution of the synthetic images matches the distribution of the real ones largely, pointing at high realism and diversity. Moreover, even though the real lesions had some outliers (pointed by the arrow as an example and were located on the negative class side), DCGAN learned the main distribution giving very few synthetic outliers (the circles on the side of the triangles).

\subsection{Radiologists assessment}
The accuracy and the ROC curve of the observer study were computed. Accuracies were reported to be $48\%$ and $61\%$ for the observers 1 and 2, respectively. Moreover, $0.56$ and $ 0.45$ Area Under the ROC curve (AUC) for observer 1 and 2, respectively, were reported. Those results suggest that the generated images appeared anatomically-plausible and were hard to be distinguished from real ones even for specialists (AUC's around random classifier performance). 

\subsection{Qualitative results}
\begin{figure}[h]
\centering
\includegraphics[scale=1]{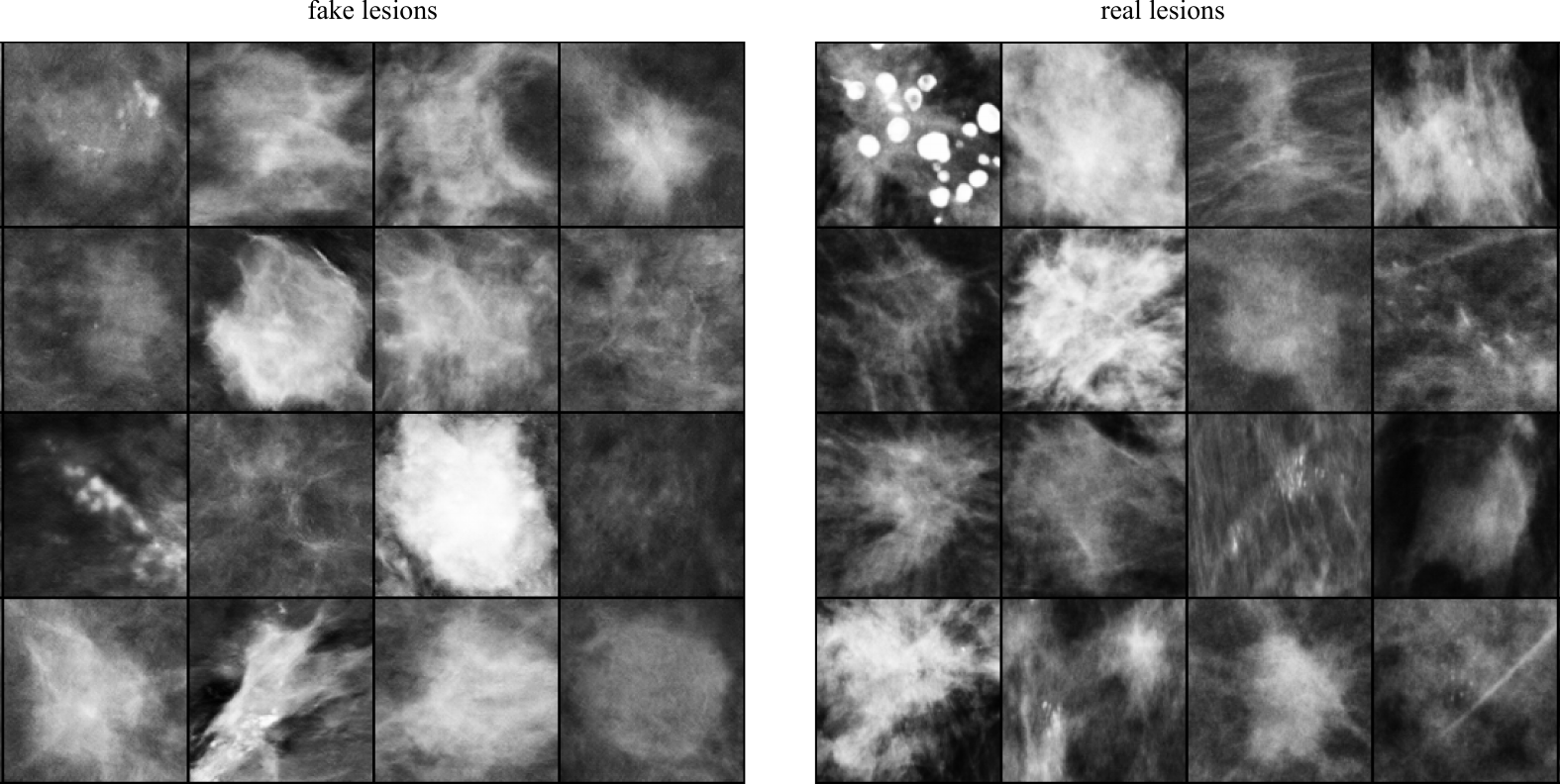}
\caption{Real (right) and fake (left) mammographic lesions.}
\label{fig:real_or_fake}
\end{figure}

A sample of 16 synthetic breast lesions (mass and microcalcification) is shown alongside a similar-size sample of real lesions. Fig. \ref{fig:real_or_fake} shows that the DCGAN could generate mammographic patches that look similar to real ones. Additionally, diversity can be identified by the different shapes and types of the generated lesions (mass, microcalcification, or mass with microcalcification).

\section{Conclusions}
Application-driven evaluations do not necessarily guarantee plausibly-looking synthetic images. Consequently, those evaluations must be accompanied by studies that care about observers assessments and feature space distribution. Here, we conducted an observers study and analysed the distribution of the real and the generated patches to strengthen our argument in our previous application-driven work. \cite{DCGAN_mammo_ours}

\bibliography{iwbi}
\bibliographystyle{spiebib} 
\end{document}